\documentclass[twocolumn,aps,floats,nofootinbib,prd,psfig,amsmath,amssymb,secnumarabic]{revtex4}
\usepackage{graphicx, epsfig, natbib}
\usepackage[nice]{nicefrac}

\newcommand{\etal}{{\sl et al.}}

\begin{document}

\title{Lack of initial conditions dependence of fluctuations in
warm inflation}
\author{Arjun Berera$^1$\footnote{\texttt{ab@ph.ed.ac.uk}}, Lisa M. H.
Hall$^2$\footnote{\texttt{lisa.hall@sheffield.ac.uk}}, Ian G.
Moss$^3$\footnote{\texttt{ian.moss@ncl.ac.uk}}, Hiranya V.
Peiris$^4$\footnote{\texttt{hiranya@ast.cam.ac.uk}}} 
\affiliation{$^1$School of Physics, University of Edinburgh, Edinburgh, EH9 3JZ \\
$^2$Department of Applied Mathematics, The University of Sheffield, Sheffield S3 7RH \\
$^3$School of Maths and Statistics, The University of Newcastle upon Tyne, Newcastle upon Tyne, NE1
7RU \\
$^4$Institute of Astronomy, University of Cambridge, Cambridge CB3 0HA}

\begin{abstract}
Warm inflation dynamics is fundamentally based on a system-reservoir
configuration in which the dynamics is dictated by a fluctuation-dissipation relation.  Recent work
by Cerioni et. al. \cite{Cerioni:2008gs}
examined dissipative
dynamics with no associated fluctuation component.
Their results, which are heavily dependent on initial conditions,
are at odds with results in the warm inflation
literature, especially the
spectral indices.
The inconsistency of their formalism is outlined here
and it is shown how this would lead to their erroneous
conclusions.  It is then shown how,
following the correct dynamical equations of warm inflation,
the results in the literature are correct.

\end{abstract}

\maketitle

The basic equation governing the evolution of inflaton modes on sub-horizon scales
in warm inflation is of the Langevin 
form \cite{Berera:1995wh,Berera:1999ws,hmb}
\begin{equation}
{\ddot {\delta \phi}} ({\bf k},t)
+ (3H + \Upsilon) {\dot {\delta \phi}}({\bf k},t)
+ ({\bf k}^2 a^{-2} + m^2) \delta \phi({\bf k}, t) = \xi({\bf k},t) ,
\label{langwi}
\end{equation}
where $\delta \phi ({\bf k},t)$ are the comoving modes of the inflaton
field $\phi({\bf x},t) = \varphi(t) + \delta \phi({\bf x},t)$, 
and $\varphi(t)$ is
the background inflaton field, specified at cosmic time $t$.
This equation emerges from the fundamental laws
of quantum dynamics \cite{qm,ab2}, accounting for expanding
spacetime.
In particular, the process of dissipation in a system is
indicative that there are other degrees of freedom interacting
with it, but that these have been integrated out, leaving
an effective evolution equation for the system.
These integrated degrees of freedom, usually called the reservoir,
will also be in some state of motion of their own.
Basic laws of dynamics imply that the interaction
between the system and reservoir will impose upon the system
dissipation effects due to energy exchange,
and forcing effects due to the motion of the reservoir degrees
of freedom.  For a given system-reservoir configuration,
with some specified state of the reservoir,
these two effects are related by a fluctuation-dissipation
theorem \cite{fd1}.  For example, the case commonly studied
in warm inflation is where the reservoir is in thermal
equilibrium.  In that case the force term in
Eq. (\ref{langwi}) is stochastic, and the fluctuation-dissipation relation
is uniquely determined to be \cite{Berera:1995wh,Berera:1999ws,hmb}
\begin{equation}
\langle \xi({\bf k}, t) \xi({\bf k}', t') \rangle =
2 (3H + \Upsilon) T a^{-3} (2\pi)^3 \delta^3({\bf k} - {\bf k}')
\delta(t-t') .
\end{equation}

The Langevin dynamics and the associated fluctuation-dissipation
relation always follow in system-reservoir configurations by
the basic quantum dynamical laws. This applies to quantum mechanical particle systems
as well as quantum field theory systems \cite{gr,bmr}.  However the dissipation
term in such equations can in general be nonlocal in space
and time, with Eq.~(\ref{langwi}) being an example of the
simplest case.

One important consequence of Eq.~(\ref{langwi}) is that the evolution
of the system, described here by the inflaton modes, is constantly affected
by the stochastic force term through the underlying dynamics of energy exchange between the inflaton
modes and the other degrees of freedom. Thus in
general, initial condition dependence of
the inflaton modes is completely wiped-out within
a short transient period, and the subsequent evolution of the inflaton modes
is governed completely by the stochastic force term.

Recently Cerioni \etal ~\cite{Cerioni:2008gs} examined the evolution
of perturbations in warm inflation under the conditions
of a thermalised statistical state.  In their formulation
they retained a dissipative term, but dropped the stochastic
force term.  This formulation is inconsistent with the
laws of physics.  Referring to our discussion above,
they have accounted for the energy exchange between system
and reservoir, but completely ignored the effect the motion
of the reservoir degrees would have on the system
during the course of the energy exchange.  This is
not consistent.

The main conclusion of Cerioni \etal~(that the spectrum from dissipative systems can be very red)
relates to the {\it homogeneous solution} of Eq.~(\ref{langwi}) 
(i.e. ignoring the stochastic source term), which is given as
\begin{equation}
\delta\phi_k = A H_\nu^{(1)} \left(-k \eta\right) a^{-3(1+r)/2}
\label{Hankelsol}
\end{equation}
where $\eta=-1/(aH)$ is the conformal time,  $A$ is a normalisation constant, $r=\Upsilon/(3H)$ and
\begin{equation}
\nu^{2}=\frac94 \left(1+r\right)^{2}-\frac{m^{2}}{H^{2}}.
\end{equation}
While we have argued that the stochastic force term is physically important when considering
dissipative terms, we will now consider the homogeneous modes for comparison with Cerioni \etal~We
will thereby show that the previous results from warm inflation are justified.

An often used approximation scheme exists whereby we match the sub-horizon scale scalar field
fluctuations to super-horizon scale curvature fluctuations at horizon crossing, i.e. when $k=aH$.
The resulting contribution to the power spectrum would be
\begin{equation}
{\cal P_R}=k^3
\left(\frac{H}{\dot\phi}\right)^{2}\vert{\delta\phi}_{k}\vert^{2},
\end{equation}
using the definition of the power spectrum from ref. \cite{hmb}.
In the massless case,
\begin{equation}
{\cal P_R} =A^2|H_\nu^{(1)}(1)|^2   
\frac{H^{5+3r}}{\dot\phi^{2}}k^{-3r},
\label{finelli}
\end{equation} 
which returns the standard quantum amplitude ${\cal P_R} = H^{4}/\dot\phi^{2}$, up to a
numerical factor, when $r=0$ and $A^2=\pi/(4H)$. Eq. (\ref{finelli}) is a simplified version of the
result found by Cerioni \etal, in which the spectrum is scale dependent: $n_{s}-1\simeq-3r$. 
However, as we will consider now, the amplitude of these fluctuations would be negligible compared
to the thermal fluctuations.

Warm inflation is defined as the regime in which $T > H$ and
where the dissipative coefficient is either large, $r > 1$, called
the strong dissipative regime or small, $r \leq 1$, called
the weak dissipative regime \cite{Berera:1999ws}.
Taking these two limits of Eq. (\ref{finelli}), 
for small and large dissipation strength $r$, gives,
\begin{eqnarray}
\mathop {\lim }\limits_{r \ll 1 }
{\cal P_R} &\sim& {4 A^2\over \pi} \frac{H^{5}}{\dot\phi^{2}}a^{-3r},
\\
\mathop {\lim }\limits_{r \gg 1 }
{\cal P_R}& \sim&{A^22^{2\nu}\Gamma(\nu)^2\over \pi^2}
 \frac{H^{4}}{\dot\phi^{2}}  a^{-3r}.
\end{eqnarray}
Although these fluctuations appear to have $n_{s}-1\simeq-3r$, in both cases the amplitude is
suppressed by a factor $a^{-3r}$ compared to the amplitude at the onset of inflation.

In the strong limit, the homogeneous solution decays rapidly and the scale dependent spectral index
becomes irrelevant.  In this case, the thermal fluctuations, ignored by Cerioni \etal, should
easily dominate.  These thermal, stochastically-produced fluctuations have already been calculated
fully~\cite{Berera:1999ws,hmb}, where it was shown that the spectrum is nearly scale-invariant, with
almost constant amplitude
\begin{equation}
{\cal P}_{\cal R}={\sqrt{\pi}\over 2}{H^2\over \dot\phi^2}(H\Upsilon)^{1/2}T
\end{equation}  
The metric perturbations were fully accounted for in this latter calculation.  Generally it can be
shown that the metric perturbations do not contribute to the fluctuation
amplitude at leading order in the slow roll parameters~\cite{Moss:2007cv}.

In the weak limit, the thermal fluctuations are still 
the dominant effect when $T\gg H$ \cite{Berera:1999ws}. The
amplitude of thermal fluctuations is constant to leading order in the slow roll parameters and can
be found using the methods described in ref.
\cite{Moss:2007cv},
\begin{equation}
{\cal P}_{\cal R}=2.29{H^3\over \dot\phi^2}T.
\end{equation}
Note that an estimate valid for all values of $r$ can be constructed by replacing $\Upsilon$ in the
result for the strong limit by $\Upsilon+3H$.

In ref. \cite{Hall:2007qw}, the authors make a crude estimate of the effect of a damping term if the
fluctuations are given by the standard super-cooled quantum limit. It is intended to investigate
the phenomenological effect on the supercooled limit of a small damping term. This estimate was
never intended (nor presented) as a full calculation of the quantum limit for dissipative
inflation, and is only valid when $r \ll 1$.

As has been argued, in warm inflation the dependence on initial conditions is very weak due to the
exponential decay of the homogeneous modes, and this is missed in the treatment of Cerioni \etal~. 
If the stochastic source term is ignored then there are serious problems with the result given by
Eq. (\ref{finelli}). These concern the choice of normalisation factor $A$. One would normally use
the natural scalar product to normalise the modes and determine $A$. However, in the dissipative
case the scalar product is no longer conserved and this normalisation makes no sense. This leads to
many problems. For example the short range power spectrum would be
\begin{equation}
{\cal P}_{\cal R}\sim A k^2a^{-2-3r},
\end{equation}
whereas on very short scales the curvature of the universe should play no role and the power
spectrum should only depend on the physical momentum $k/a$. These
problems are resolved by a correct treatment of the dissipative system which requires either using
a density matrix formalism or the Langevin equation \cite{qm,ab2}. 

We have laid stress on the fact that the homogeneous solutions to the fluctuation equation for warm
inflation decay exponentially, weakening the effect of initial conditions on the fluctuation
amplitude. The fluctuation amplitude depends on a balance between the fluctuations induced by the
radiation through a source term and the tendency to decay due to the expansion of the universe.

\acknowledgments 
We would like to thank Alessandro Cerioni, Fabio Finelli and Alessandro Gruppso for discussing their
preprint with us.


\end{document}